\documentclass[aps, reprint, prb, superscriptaddress, floatfix]{revtex4-2}
\usepackage{graphicx}
\usepackage{amssymb}
\usepackage{amsmath}
\usepackage{mathrsfs}
\usepackage[colorlinks,linkcolor=blue,anchorcolor=blue,citecolor=blue,urlcolor=blue]{hyperref}
\usepackage{bm}
\usepackage{multirow}
\usepackage{xcolor}
\usepackage{lineno}

\begin{document}
\title{Full tunability and quantum coherent dynamics of a driven multilevel system}
\author{Yuan Zhou}
\affiliation{CAS Key Laboratory of Quantum Information, University of Science and Technology of China, Hefei, Anhui 230026, China}
\affiliation{CAS Center for Excellence and Synergetic Innovation Center in Quantum Information and Quantum Physics, University of Science and Technology of China, Hefei, Anhui 230026, China}

\author{Sisi Gu}
\affiliation{CAS Key Laboratory of Quantum Information, University of Science and Technology of China, Hefei, Anhui 230026, China}
\affiliation{CAS Center for Excellence and Synergetic Innovation Center in Quantum Information and Quantum Physics, University of Science and Technology of China, Hefei, Anhui 230026, China}

\author{Ke Wang}
\affiliation{CAS Key Laboratory of Quantum Information, University of Science and Technology of China, Hefei, Anhui 230026, China}
\affiliation{CAS Center for Excellence and Synergetic Innovation Center in Quantum Information and Quantum Physics, University of Science and Technology of China, Hefei, Anhui 230026, China}

\author{Gang Cao}
\affiliation{CAS Key Laboratory of Quantum Information, University of Science and Technology of China, Hefei, Anhui 230026, China}
\affiliation{CAS Center for Excellence and Synergetic Innovation Center in Quantum Information and Quantum Physics, University of Science and Technology of China, Hefei, Anhui 230026, China}
\affiliation{Hefei National Laboratory, University of Science and Technology of China, Hefei 230088, China}

\author{Xuedong Hu}
\affiliation{Department of Physics, University at Buffalo, SUNY, Buffalo, New York 14260, USA}

\author{Ming Gong}
\email[]{gongm@ustc.edu.cn}
\affiliation{CAS Key Laboratory of Quantum Information, University of Science and Technology of China, Hefei, Anhui 230026, China}
\affiliation{CAS Center for Excellence and Synergetic Innovation Center in Quantum Information and Quantum Physics, University of Science and Technology of China, Hefei, Anhui 230026, China}
\affiliation{Hefei National Laboratory, University of Science and Technology of China, Hefei 230088, China}

\author{Hai-Ou Li}
\email[]{haiouli@ustc.edu.cn}
\affiliation{CAS Key Laboratory of Quantum Information, University of Science and Technology of China, Hefei, Anhui 230026, China}
\affiliation{CAS Center for Excellence and Synergetic Innovation Center in Quantum Information and Quantum Physics, University of Science and Technology of China, Hefei, Anhui 230026, China}
\affiliation{Hefei National Laboratory, University of Science and Technology of China, Hefei 230088, China}

\author{Guo-Ping Guo}
\affiliation{CAS Key Laboratory of Quantum Information, University of Science and Technology of China, Hefei, Anhui 230026, China}
\affiliation{CAS Center for Excellence and Synergetic Innovation Center in Quantum Information and Quantum Physics, University of Science and Technology of China, Hefei, Anhui 230026, China}
\affiliation{Hefei National Laboratory, University of Science and Technology of China, Hefei 230088, China}
\affiliation{Origin Quantum Computing Company Limited, Hefei, Anhui 230026, China}

\date{\today}

\begin{abstract}
Tunability of an artificial quantum system is crucial to its capability to process quantum information.
However, tunability usually poses significant demand on the design and fabrication of a device. In this work, we demonstrate that Floquet engineering based on longitudinal driving provides distinct possibilities in enhancing the tunability of a quantum system without needing additional resources. In particular, we study a multilevel model based on gate-defined double quantum dots, where coherent interference occurs when the system is driven longitudinally.  We develop an effective model to describe the driven dynamics of this multilevel system, and show that it is highly tunable via the driving field.  We then illustrate the versatility and rich physics of a driven multilevel system by exploring phenomena such as driving modulation of resonances, adiabatic state transfer, and dark state.  In the context of qubit control, we propose noise-resistant quantum gates based on adiabatic passage.  The theoretical consideration we present here is rather general, and is in principle valid for other multilevel quantum systems.

\end{abstract}

\maketitle
\section{Introduction}
Recent developments in quantum coherent devices has pushed the state-of-the-art ever closer to the threshold of fault-tolerant quantum computing \cite{Arute2019, Figgatt2019, Egan2021, Wu2021, Hendrickx2021, Adam2022, Noiri2022, Xue2022, Madzik2022, Camenzind2022}.
Tunability of a quantum system is a crucial ingredient and an enabler to its utility in quantum information processing.
While more tunable usually means additional elements in the device design, and therefore more complex to fabricate, Floquet engineering via a periodic drive \cite{Bukov2015, Oka2019, Rudner2020, Weitenberg2021} could potentially help solve this dilemma.  Driving has been employed to tailor quantum systems \cite{Jimenez2015, Clark2019, Wang2019}, and leads to interesting novel phenomena such as Floquet topological insulators \cite{Rudner2020, Weitenberg2021}.  Applied strategically, Floquet engineering would generate significant tunability while avoiding complex designs: under a periodic drive, a qubit is dressed by the photons, and the driving field amplitude, frequency, and phase would all provide possible new tuning knobs for the dressed qubit \cite{Laucht2017, Nakonechnyi2021}.

Atoms are the prototypical transversely driven systems, where the bare atomic spectrum provides a reference frame on which the atom-photon coupling (i.e. driving) appears on the off-diagonal of the total Hamiltonian and generates transitions between different atomic levels.  As such, when the driving frequency is near resonance with a particular transition, the driven dynamics is well described by a two-level model. In contrast, in many artificial quantum systems, such as superconducting nanocircuits and gate-defined double quantum dots (DQD), the coupling to the external alternating field could appear on the diagonal of a non-driven system and acts as a longitudinal drive, and could introduce rich physics since multiple states may be involved, with both weak and strong longitudinal drive easily achievable in practice. 

A multilevel system with a weak or moderate driving amplitude is most often studied in the far-detuned regime (away from any level anti-crossings), such as electric dipole spin resonance (EDSR) where driving amplitude is typically much less than the energy splitting, and effects of the intermediate states are treated perturbatively \cite{Zajac2018, Takeda2020, Hendrickx2020}. An example of a strongly longitudinally driven system is the Landau-Zener-St\"uckelberg (LZS) interference \cite{Shevchenko2010, Silveri2017, Ivakhnenko2022} in two-level systems \cite{Oliver2005, Ashhab2007, Oliver2009, Cao2013, Stehlik2012}.  In this context a multilevel system is intriguing due to possible higher-order interference among multiple paths of evolution. However, some multilevel studies are still dominated by two-level dynamics, so that they can be well characterized by effective two-level models \cite{Sun2010, Mi2018, Forster2014, Shevchenko2018}; while in other cases the results have been interpreted by numerical simulations \cite{Bogan2018}, or by introducing additional assumptions \cite{Danon2014}, which tend to obscure the original multilevel physics.  An effective model based on high-frequency expansion has been developed \cite{Rahav2003, Goldman2014, Eckardt2015}, though we would show that it fails to capture some interesting multilevel dynamics studied here. 
In short, understanding of a longitudinally driven multilevel system remains incomplete at present \cite{Ivakhnenko2022}.

Quantum dots provide a viable platform for quantum computation and quantum simulation, having advantages such as high degree of integration, all-electrical control, and potentially higher operating temperatures \cite{Zhang2019, Petit2020, Yang2020, Camenzind2022}.  A double quantum dot can be conveniently driven longitudinally when an alternating voltage is applied to top gates that control the interdot detuning.  Indeed, with a strong longitudinal drive of broad bandwidth, LZS interference has been widely observed, and applied to qubit manipulation \cite{Cao2013}, dephasing characterization \cite{Forster2014}, spectroscopy \cite{Stehlik2012}, {\it{etc}}. With precise state control and measurement available for electrons in QDs, DQDs provide an ideal platform for exploring longitudinally driven multilevel systems, and will be the nominal system we explore in this work.

Here we study a prototypical multilevel quantum system under a longitudinal drive by first deriving a generic effective model.  Specifically, we perform a unitary transformation to move the time-dependence to the off-diagonal of the Hamiltonian, expand it in the Floquet basis, and then perform a Schrieffer-Wolff (SW) transformation into the First Brillouin Zone (FBZ) \cite{Winkler2003}. 
The resulting effective Hamiltonian in the FBZ captures quite accurately the multilevel driven dynamics.  It does not rely on approximations such as strong driving or strong dissipation, thus is highly versatile and allows us to investigate the driven system under a variety of conditions.
Based on this model, we demonstrate possible observation under longitudinal drive of several celebrated quantum coherent phenomena that have been observed in atomic and other systems via transverse driving, such as modulation of resonances by driving \cite{Autler1955, Xu2007, Sillanpaa2009}, electromagnetically induced transparency \cite{Gray1978, Boller1991, scully1997, Wei1999, Eisaman2005, Xu2008, Boyd2008}, and adiabatic state transfer \cite{Kuklinski1989, Takekoshi2014, Kumar2016, Vitanov2017}.
We investigate these phenomena both numerically and theoretically, with focus on the distinct features due to longitudinal driving.
Lastly, we explore applications of the high-order LZS interference in the context of qubit control in a two-spin system, therefore broadening the prospect of QDs for future quantum coherent devices.

\begin{figure}
    \centering
    \includegraphics[width=8.3 cm]{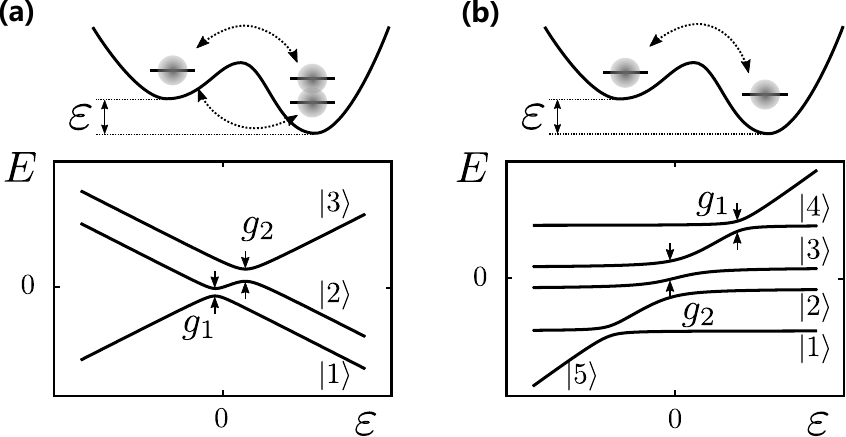}
    \caption{{\bf Gate-defined QDs and their lowest energy levels}.
	(a) Realization of the three-level model based on a hybrid qubit with three spins.
	(b) Realization of the five-level model based on the singlet-triplet qubit system.
	In these systems, the off-diagonal couplings $g_i$ between different states are
	realized based on tunneling through the barrier and spin-orbit coupling, while
	the energies in the diagonal terms can be tuned by the detuning
	between the two quantum dots, denoted by $\varepsilon$.}
	\label{fig-setup}
\end{figure}

\section{Theoretical Model}

In this section we discuss the multilevel physical systems as shown in Fig.~\ref{fig-setup} (a) and (b), then derive a generic effective model Hamiltonian to describe these systems under longitudinal drive.

A three-level system is the minimal prototype to demonstrate higher order quantum interference.  One example is the three-spin hybrid qubit \cite{Shi2012, Kim2014, Shi2014, Cao2016} in an asymmetric Si DQD, as shown in Fig.~\ref{fig-setup} (a).  In the subspace with total spin $S=1/2$ and $S_z=-1/2$, the lowest three levels can be isolated from the excited manifolds, yielding the following basis \cite{Shi2012}
\begin{eqnarray}
	&& |1\rangle = |\downarrow,S\rangle, \quad |2\rangle = \sqrt{{\frac{1}{3}}}|\downarrow,T_0\rangle - \sqrt{{\frac{2}{3}}}
|\uparrow,T_-\rangle, \nonumber \\
	&& |3\rangle = |S,\downarrow\rangle, \label{eq-basis1}
\end{eqnarray}
where $|S\rangle = (|\uparrow\downarrow\rangle - |\downarrow\uparrow\rangle)/\sqrt{2}$,
$|T_0\rangle = (|\uparrow\downarrow\rangle + |\downarrow\uparrow\rangle)/\sqrt{2}$ and
$|T_-\rangle = |\downarrow\downarrow\rangle$ in $|1\rangle$ and $|2\rangle$ are the singlet and triplet states in the right dot, while $|S\rangle$ in $|3\rangle$ is in the left dot, giving it a different charge occupation and detuning dependence (double occupation in the left instead of right dot) from $|1\rangle$ and $|2\rangle$.

The Hamiltonian governing the three-level system in Fig.~\ref{fig-setup} (a) takes the form
\begin{equation}\label{eq-H_ThLS}
    H(t) =
    \begin{pmatrix}
        D_1 & 0 & g_1\\
        0 & D_2 & g_2\\
        g_1 & g_2 & D_3(t)
    \end{pmatrix},
\end{equation}
where we have set $\hbar =1$, and the unit for energy is gigacycle (Gc) throughout the paper.  The driving of this system is through the interdot detuning and is thus longitudinal within the current basis, 
$$D_3(t) = \varepsilon -  A\cos(\omega t) \,.$$  
The couplings $g_1$ and $g_2$ are given by the inter-dot tunneling, and are set as constants in this study.

A five-level spectrum similar to Fig.~\ref{fig-setup} (b) has been explored in a two-electron singlet-triplet (ST) system in a DQD \cite{Petta2005, Pioro-Ladriere2008, Noiri2022, Xue2022}.
The relevant basis states are
\begin{eqnarray}
&& |1\rangle = |\uparrow,\uparrow\rangle, \quad |2\rangle = |\uparrow,\downarrow\rangle, \quad |3\rangle = |\downarrow,\uparrow\rangle, \nonumber \\
&& |4\rangle = |\downarrow,\downarrow\rangle, \quad |5\rangle = |S,~0\rangle.
\label{eq-basis2}
\end{eqnarray}
Here states $|1\rangle$ to $|4\rangle$ have one electron in each dot, while $|5\rangle$ has both electrons in the left dot in Fig.~\ref{fig-setup}(b).  In such an ST system, the coupling denoted by $g_2$ (couplings between $|2\rangle$, $|3\rangle$ and $|5\rangle$) is given by spin-conserved inter-dot tunneling, while $g_1$ (couplings between $|1\rangle$, $|4\rangle$ and $|5\rangle$) represents spin-flip tunneling enabled by spin-orbit interaction. Usually for conduction electrons $g_1 \ll g_2$. 
In addition, state $|3\rangle$ in Eq.~(\ref{eq-basis1}) and $|5\rangle$ in Eq.~(\ref{eq-basis2}) respond differently to the interdot detuning $\varepsilon$ due to their different charge configurations as is shown in Fig.~\ref{fig-setup}, and consequently can be detected electrically.  Hereafter they will be termed as the shuttle state.

In a longitudinally driven system, the strong and weak driving limits are defined based on the comparison of driving amplitude and splittings of anticrossings \cite{Ashhab2007}.  Compared to a two-level model, a multilevel system usually has multiple anticrossings.  As such we define strong driving in our multilevel models based on whether a driven system passes through multiple anticrossings repeatedly. In the following studies, we are particularly interested in the regime where the detuning difference between anticrossings is of the same order of magnitude as $\omega$, so that the strong driving condition is given by $(A-\varepsilon) \gg \omega$, although we also study situations where weaker driving is sufficient.

\subsection{The effective Hamiltonian}\label{sec-Heff}

To solve for the dynamics governed by the Hamiltonian given in Eq.~(\ref{eq-H_ThLS}), we perform unitary transformations to shift the time dependence into the basis states, and search for a time-independent effective Hamiltonian.  Specifically, we first eliminate the time dependence in the diagonal terms of the Hamiltonian by going to a rotating frame through the rotation $\mathcal{U} = \text{diag} (e^{i (D_2+n\omega) t}, e^{i D_2 t}, e^{i\int_0^td\tau [D_2+N\omega-A\cos(\omega\tau)]})$, where $n$, $N$ are integers. The Hamiltonian in this rotating frame is given by $\tilde{H}(t) = \mathcal{U}^\dagger H(t) \mathcal{U} - i \mathcal{U}^\dagger \partial_t \mathcal{U}$. Note that $n\omega$ and $N\omega$ are two frequency offsets, which ensure time-translation symmetry for the Hamiltonian in the new basis, $\tilde{H}(t) = \tilde{H}(t+T)$, and minimize its diagonal terms.  We then perform the Floquet expansion \cite{Shirley1965} on $\tilde{H}$ to separate the different multi-photon processes, followed by a SW transformation back into the FBZ (where the diagonal terms are minimized) in the frequency domain \cite{Winkler2003}.  The details of this procedure are presented in the Appendix~\ref{appdx-Heff}.

The effective time-independent Hamiltonian of our driven three-level system takes the form
\begin{equation}
	\label{eq-H_eff}
	H_{\text{eff}} =
    \begin{pmatrix}
        D_{12} - n \omega +\Delta' 		& \xi & g_1 J_{N-n} \\
        \xi^* 	& \Delta & g_2 J_{N} \\
        g_1^* J_{N-n} 	& g_2^* J_{N} & -\delta-\Delta - \Delta'
    \end{pmatrix},
\end{equation}
where $D_{12} = D_1 - D_2$, $\delta \equiv D_2-\varepsilon+N\omega$, and $J_k = J_k(\frac{A}{\omega})$ is the first-kind Bessel function of the $k$-th order. $\Delta'$, $\Delta$, and $\xi$ are corrections from the higher harmonics via the second-order SW transformation.  Hereafter we also denote $D_i-\varepsilon$ as $D_{i3}$ (i.e. $D_{i3}$ does not include the driving field, and $D_{3i} = -D_{i3}$).  
The basis states for this effective Hamiltonian are rotated from the original ones, and are defined by a diagonal matrix $q$ and a kick operator $K$ as $|\psi_\text{eff}\rangle = e^{-iqt}e^{-iK(t)}|\psi\rangle$ (See Appendix \ref{appdx-comp}).

$H_\text{eff}$ remains accurate as long as the perturbative treatment in SW transformation is valid, which requires $g_i J_\nu(A/\omega)/|D_i-\varepsilon-m\omega| \ll 1$, with $i = 1,2$, $\nu\ne 0$, and $m\omega$ an arbitrary frequency offset.  Notice that this condition is essentially a comparison between driving frequency detuning and the tunnel coupling (modified by a Bessel function bound by 1).  Driving amplitude only shows up as a variable in the Bessel function, and does not pose a strong restriction on the validity condition here, so that our effective Hamiltonian is applicable in both weak and strong driving regimes.
	
When $D_{12} = n\omega$ and $\varepsilon-D_2 =  N\omega$, the effective Hamiltonian $H_{\rm eff}$ is equivalent to the one obtained by the high-frequency expansion \cite{Goldman2014}. However, Eq.~(\ref{eq-H_eff}) is valid in a much broader parameter regime because of the more adaptable transformations involved in its derivation. For example, the difference can be obvious when $\varepsilon -D_2=\omega/2$, as we discuss in Appendix \ref{appdx-Heff} and \ref{appdx-comp}. 

The effective Hamiltonian shows that the original single longitudinal drive has made the off-diagonal couplings individually tunable via the Bessel functions.  This wide tunability is highly desirable in a hybrid qubit (and the singlet-triplet qubit we discuss below) since it is challenging to tune $g_1$ and $g_2$ individually in practice.  Furthermore, a synthetic interaction labeled by $\xi$ is now generated, coupling the dressed qubit states $|1\rangle_\text{eff}$ and $|2\rangle_\text{eff}$.

The same approach can be applied to derive an effective model for a five-level system with the ST energy spectrum of Fig.~\ref{fig-setup}(b). 
In fact, a similar three-level model can also be obtained from the five-level model when the system is near resonance: For example, when $D_{12}\approx n\omega$, we can take $\{|1\rangle,|2\rangle,|5\rangle\}$ into consideration to construct an effective Hamiltonian while excluding the other two states, which contribute little to the dynamics supposing that the spectrum is well addressable and the system is initialized to $|1\rangle$ or $|2\rangle$ (See Appendix~\ref{appdx-5vs3}). As we would show later, the three-level model agrees well with the numerical simulation based on the full five-level model Hamiltonian. Thus, we expect that similar features, such as driving induced shifts in resonances, odd-even effect, and dark state, as we will discuss in the next section, are all observable in the ST system as well.

\section{Results}

In this section we study the consequences of longitudinal driving in a few different multilevel contexts. In subsection \ref{sec-Resonances} we focus on the modulation of resonance spectrum with a moderate driving strength $(A-\varepsilon) \sim \omega$. In addition, we discuss modulation of resonance intensities by driving, especially focusing on the odd-even effect that appears when approaching the strong driving limit. In subsection \ref{sec-dark_state} we discuss a dark state observed at the strong driving limit when $(A-\varepsilon) \gg \omega$.
Subsection \ref{sec-qgate} is devoted to a quantum gate based on synthetic couplings induced by the longitudinal driving, which lies in the regime of intermediate driving strength $A-\varepsilon \sim \omega$. And in subsection \ref{sec-AST} we present a proposal to realize coherent state transfer by increasing driving amplitude adiabatically.

\subsection{Resonances, Driving Modulation of Splittings, and Odd-Even Effect}\label{sec-Resonances}

\begin{figure}
	\centering
	\includegraphics[width=8.3 cm]{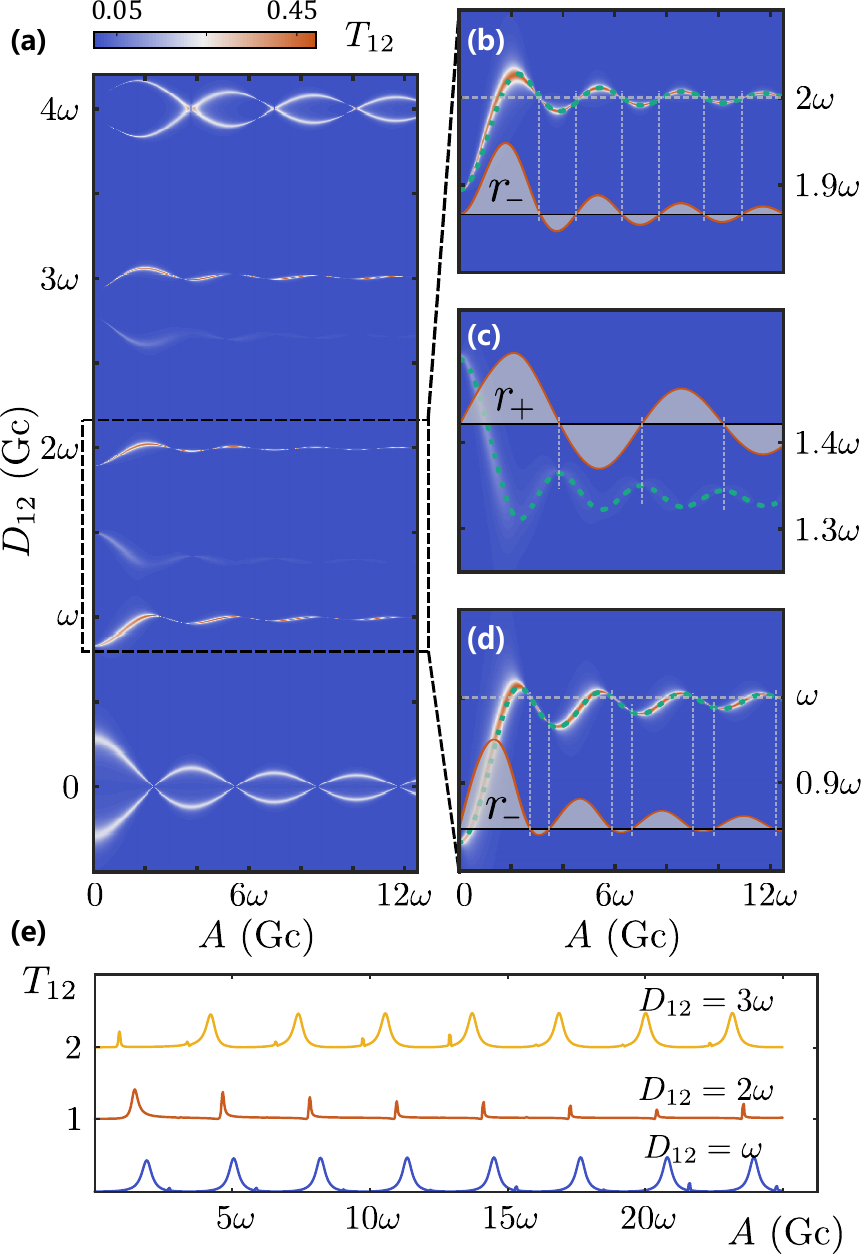}
	\caption{{\bf The transition amplitude of the driven three-level model}.
		(a) Time-averaged transition amplitude $T_{12}$ as a function of driving amplitude $A$ and energy spacing $D_{12}$ with parameters $g_1=0.1$ Gc, $g_2= 1$ Gc, $\omega = 4$ Gc, $D_1 = -3D_2$ and $\varepsilon=0$. Significant transition is
		observed at the resonant condition of Eq.~\ref{eq-resonance}.
		Zoom-in of $D_{12}/\omega \in [0.8, 2.2]$ are placed in (b)-(d), in which the green dotted lines represent the resonance indicated by $H_\text{eff}$, and the shaded functions are the effective couplings of $r_\pm$, zero points of which yield the break points.
        Breakpoints in (b) are well aligned to $D_{12}=2\omega$ while those in (d) are not aligned to $D_{12}=\omega$, which is the evidence of odd-even effect.
        (e) Cross-sections of (a) in $D_{12}=\omega$, $2\omega$ and $3\omega$. Signals are shifted for clarity. For $D_{12}=2\omega$, the transition is suppressed, and the suppression gets severe as driving amplitude increases.
        }
	\label{fig-cdt}
\end{figure}

We first investigate resonant conditions for transitions and modulation of effective couplings (thus the intensities of the resonances) using both our effective Hamiltonian and direct numerical simulation with the original time-dependent Hamiltonian.  In Fig.~\ref{fig-cdt} (a) we present the long-time averaged transition amplitude $T_{12}$ between $|1\rangle$ and $|2\rangle$ defined by $\lim_{t\to\infty}\int_0^t|\langle 1|U(\tau)|2\rangle|^2d\tau$, where $U(t)$ is the time evolution operator.(see Appendix \ref{appdx-Heff} for details) as functions of driving amplitude $A$ and state splitting $D_{12}$.
The numerical results here are obtained using the original $H(t)$, with $g_1=0.1$ Gc, $g_2=1$ Gc and $\omega=4$ Gc.  

At the weak driving limit ($A \ll \omega$ such that $A/\omega \rightarrow 0$), we can see clearly a sequence of resonances, with the most prominent resonances around $D_{12} \sim \pm \frac{1}{3} \omega$.  The higher harmonic resonances fade away as $A \rightarrow 0$ because they are not accessible by weak driving.  Indeed, when $A = 0$, the system is not driven (so that Eq.~(\ref{eq-H_ThLS}) is time-independent) and can be solved analytically.  The finite tunnel coupling $g_2$ mixes $|2\rangle$ and $|3\rangle$, and $g_1$ allows finite transition amplitudes from $|1\rangle$ to the mixtures of $|2\rangle$ and $|3\rangle$, corresponding to the two bright peaks on either side of $D_{12} = 0$. With our particular choices of parameters, we obtain the locations of the two peaks at $D_{12} = \pm \frac{2}{\sqrt{3}} g_2 \approx \pm 0.3 \omega$, consistent with our numerical results.

As the driving amplitude $A$ increases, the higher harmonic resonances grow in prominence, while their locations and intensity are modified by both the driving amplitude $A$ and the splitting $D_{12}$ (recall that the driving frequency $\omega$ is kept as a constant). 
The effective model Hamiltonian $H_\text{eff}$ provides a clear framework to understand the various features of Fig.~\ref{fig-cdt} (a). In particular, the resonant transition conditions are determined by the diagonal matrix elements.  For simplicity, we assume $g_1 \ll g_2$ (applicable when studying a driven ST system), so that we can neglect the effect of $\Delta'$ and obtain
\begin{equation}
	D_{12} = n\omega -\frac{\delta}{2} \pm \sqrt{\left(g_2 J_N(A/\omega)\right)^2 + \left(\Delta + \delta/2\right)^2} \,, \\
\label{eq-resonance}
\end{equation}
for the principal resonance and higher harmonics of the two sets of transitions, with $N$ and $n$ both integers.  This equation can only be solved numerically and self-consistently since $\Delta$ depends on $D_{12}$, as given in Appendix A.3. Nevertheless, the presence of $J_N(A/\omega)$ means that the positions of the resonances are dependent on the driving amplitude $A$ through the Bessel functions, as illustrated in Fig.~\ref{fig-cdt} (b)-(d), where the analytical results from the effective model are presented by the green dotted lines, in excellent agreement with the numerical simulation.

Figure \ref{fig-cdt} (a) shows that the resonances often appear in pairs, as are most clearly shown near $D_{12} \sim 0$ and $D_{12} \sim 4\omega$ (in between the pairing of resonances is not clear, because we have chosen $D_1 = -3 D_2$ in our simulations, causing two of the resonances to merge). These pairs of resonances are caused by the coupling and mixing between $|2\rangle_\text{eff}$ and $|3\rangle_\text{eff}$, so that two absorption peaks develop when considering transitions from $|1\rangle_\text{eff}$.  Indeed, if we had calculated $T_{13}$, we would have obtained the same patterns as in Fig.~\ref{fig-cdt}, as the destination states are mixtures of $|2\rangle_\text{eff}$ and $|3\rangle_\text{eff}$.  Notice that the coupling here are tunnel couplings between the bare $|2\rangle$ and $|3\rangle$ states, modified by the Bessel function $J_N(A/\omega)$, such that finite driving would lead to modulation of the effective coupling strength between the dressed states $|2\rangle_\text{eff}$ and $|3\rangle_\text{eff}$.  
The splittings in the resonant signals for each $n$-harmonic appear quite similar to the Autler-Townes Splitting (ATS) in a transversely driven system, where a resonance splits into two due to Rabi oscillations from driving \cite{Autler1955, Xu2007, Sillanpaa2009}.  However, upon closer inspection, we see that here the splittings exist even in the absence of driving as they originate from tunnel coupling between the two quantum dots. When the system is driven, the driving field does modulate the splittings in the same spirit of ATS, though here the modulation is in the form of $J_N(A/\omega)$ as is shown in Fig.~\ref{fig-cdt} (b)-(d) by the green dotted lines, because the driving here is longitudinal.  As the driving amplitude increases into the strong drive $A/\omega\gg 1$ regime, the $A$-dependence saturates and the resonance peak locations approaches $D_{12} = n\omega$, again different from the original ATS.

\begin{figure*}
    \centering
    \includegraphics[width=17.2 cm]{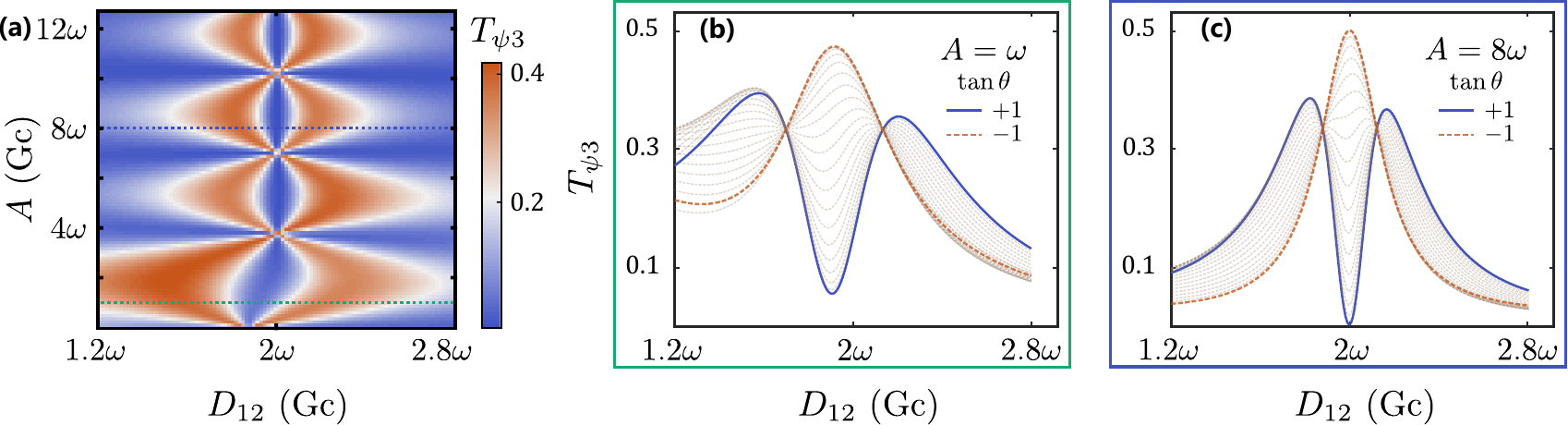}
	\caption{{\bf Dark state based on the effective three-level model}.
    (a) Averaged transition amplitude $T_{\psi3}$ as a function of
	$D_{12}$ and $A$, with $D_{13}=D_{32}$, $g_1=g_2=1$ Gc, $\omega = 4$ Gc and $\theta = \pi/4$.
    We focus on the region of $D_{12}\approx 2\omega$ with effective couplings given by $g_1J_{-1}(A/\omega)$ and $g_2J_{1}(A/\omega)$, respectively.
    As a result, $|\psi(\pi/4)\rangle$ is the dark state in case of $D_{12}=2\omega$ with ignorable $\Delta$, $\Delta'$ and $\xi$.
    (b) and (c) show the cross-sections at $A = \omega$ and $8\omega$, respectively, with different initial state by scanning the relative phase in $|\psi(\theta)\rangle$.
    When $A/\omega$ is not large enough, the dip will not approach zero due to the tiny contribution of $\Delta$, $\Delta'$ and $\xi$.}
	\label{fig-CPT}
\end{figure*}

Simulation results in Fig.~\ref{fig-cdt} also show that changing driving amplitude $A$ also modulate the intensities of the resonances.
In particular, as driving amplitude $A$ increases, we observe break points along the resonant curves where $T_{12}$ vanishes or is strongly suppressed.  To understand this feature, we examine more closely the regime where hybridization between $|2\rangle_\text{eff}$ and $|3\rangle_\text{eff}$ is strong.  In order to evaluate the transition amplitude accurately, we first diagonalize the subspace of $\{|2\rangle_\text{eff},~|3\rangle_\text{eff}\}$ as $\text{diag}(E_+, E_-)$, with the new basis denoted by $\{|+\rangle,~|-\rangle\}$.  Effective couplings between $|1\rangle_\text{eff}$ and $|\pm\rangle$ denoted by $r_\pm$ are presented in the inset in Fig.~\ref{fig-cdt} (b)-(d), showing that the break points occur when $r_\pm = 0$. In other words, the break points mean that tunneling between $|1\rangle_\text{eff}$ and $|\pm\rangle$ is forbidden, similar to the coherent destruction of tunneling (CDT) in a driven two-level system \cite{Grossmann1991, Oliver2005, Ashhab2007, Stehlik2012}, extending to a multilevel system here.

The driving-modulated resonances together with CDT leads to a striking odd-even alternating effect.
In Fig.~\ref{fig-cdt} (b) the breakpoints are well aligned to $D_{12}=2\omega$ while the distribution of the breakpoints in Fig.~\ref{fig-cdt} (d) are different and shift away from $D_{12} = \omega$.
Generally, for $D_{12} = n\omega$, the transition amplitude for even $n$'s is always suppressed, while that for odd $n$'s is not, thus we term this phenomenon as odd-even effect.
In Fig.~\ref{fig-cdt} (e), we present $T_{12}$ as a function of $A$ under the conditions of $D_{12} = \omega$, $2\omega$ and $3\omega$ for a better description of the odd-even effect.
Here the transition amplitude display discrete peaks due to the modulated effective coupling.
When resonant peaks meet CDT, the peaks are suppressed. 
For $D_{12} = 2\omega$, the peaks are narrower compared to that of $D_{12} = \omega$ and $3\omega$, indicating reduced overall absorption, and in agreement with the fact that the breakpoints are aligned to $D_{12} = 2\omega$.

The origin of the odd-even effect is the driving-induced modulation of the effective coupling in $H_\text{eff}$.
Again assuming $g_1J_{N-n} \ll g_2J_N$, for the resonances of $D_{12}=n\omega$ and $D_{13}=n\omega$ the transitions are determined by $r_-$ and $r_+$, respectively.  Assuming that $\Delta'$ is negligible, we obtain $r_-$ under the condition $D_{12}-n\omega=E_-$ and $r_+$ under the condition $D_{12}-n\omega=E_+$:
\begin{equation}
    \begin{aligned}
        r_- &\propto \xi g_2J_N - g_1J_{N-n}(\Delta-E_-),\\
        r_+ &\propto \xi g_2J_N + g_1J_{N-n}(E_+-\Delta).
    \end{aligned}
\end{equation}
By substituting in the expressions of $\Delta$ and $\xi$, and accounting for the condition $E_-=0$, which means $D_{12}=n\omega$, we find $r_-\propto \sum_{m \ne N}g_1g_2^2(J_{m-n}J_{m}J_{N}-J_{m}^2J_{N-n})/(m\omega+D_{23})$ (see Appendix \ref{appdx-odd-even}).
In the large $A/\omega$ limit (i.e. strong driving), the asymptotic behavior of the Bessel functions dictates that for even $n$, $r_-\approx 0$. Thus transitions at $D_{12}=n\omega$ for large driving amplitude $A$ is forbidden, yielding the so-called odd-even effect.
This effect was experimentally observed in a DQD by Stehlik {\it et al.} \cite{Stehlik2014}, and explained based on an assumption of strong dissipation \cite{Danon2014}.
Our results here show that even without dissipation, the odd-even effect could be observable because of the CDT in a multilevel system in the strong driving limit.
We further discuss the odd-even effect with dissipation in 
Appendix \ref{appdx-odd-even-diss}, which is consistent with the results in Ref.~\onlinecite{Stehlik2014, Danon2014}.

\subsection{Dark state in the effective Hamiltonian}\label{sec-dark_state}

In atomic physics, a dark state is a superposition of the two lower-energy states in a three-level $\Lambda$ system \cite{Gray1978}. It is decoupled from the excited state due to interference effects from transverse driving.  When a system relaxes to the dark state, it will be trapped there \cite{Gray1978, Wei1999, Eisaman2005, Xu2008}, leading to coherent population trapping.  In our longitudinally driven system, off-diagonal coupling is present in the effective Hamiltonian $H_{\text{eff}}$ among the dressed states as a result of the higher-order LZS interference.  Accordingly, when $\xi$ is negligible, we expect that a dark state can form for our system:
\begin{equation}
	|\psi\rangle \propto g_2 J_N |1\rangle_\text{eff} - g_1 J_{N-n} |2\rangle_\text{eff}.
\end{equation}
When the system is trapped in the dark state, the transition between $|\psi\rangle$ and the shuttling state $|3\rangle$ is strictly forbidden.  Considering that state $|3\rangle_\text{eff}$ has a different charge configuration from $|1\rangle_\text{eff}$ and $|2\rangle_\text{eff}$, forbidden transition to $|3\rangle_\text{eff}$ means charge cannot transfer between the two dots, leading to a blockade of current through the DQD.  In essence, coherent trapping here is caused by interference between the two paths of electron tunneling.

\begin{figure*}
    \centering
    \includegraphics[width=17.2 cm]{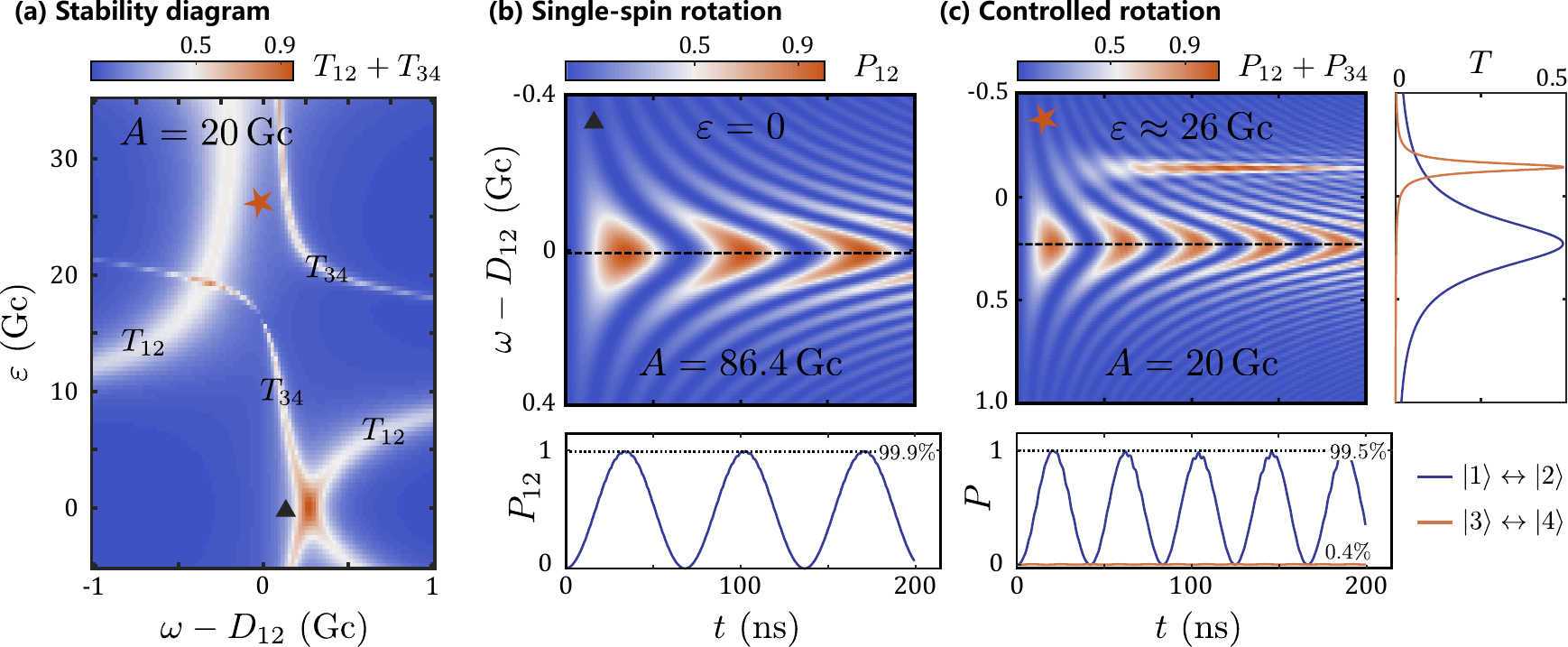}
    \caption{{\bf Quantum gate via Rabi oscillation in the ST system}.
	(a) Total transition amplitude $T_{12}+T_{34}$ versus detuning $\varepsilon$ and driving frequency.
    Parameters used in simulation are $g_1=1$ Gc, $g_2=2$ Gc, $D_1=-D_4=40$ Gc and $D_2=-D_3=10$ Gc.
    (b) Rabi chevron for single-spin rotation with suppressed leakage, with $\varepsilon=0$ and $A=86.4$ Gc.
    The resonant oscillation has a maximal amplitude over $99.9\%$.
	(c) Two-qubit conditional-rotation gate with $\varepsilon = 26$ Gc and $A = 20$ Gc, where resonant transitions for $P_{12}$ and $P_{34}$ are at different frequencies.
    }
    \label{fig-Rabi}
\end{figure*}

In Fig.~\ref{fig-CPT} we demonstrate the emergence of the dark state by plotting the average transition amplitude $T_{\psi 3}$ from a random initial state $|\psi(\theta)\rangle = \cos(\theta) |1\rangle_\text{eff} + \sin(\theta) |2\rangle_\text{eff}$ to $|3\rangle_\text{eff}$ with $g_{1,2} = 1$ Gc. In the vicinity of $D_{12}=2\omega$ and $D_{13}=\omega$, with $n=2$, $N=1$, the dark state $|\psi\rangle$ would be given by $|\psi(\pi/4)\rangle$.
Numerical simulation of averaged transition amplitude $T_{\psi 3}$ is shown in Fig.~\ref{fig-CPT} (a), in which the suppression of transition amplitude is observed around $D_{12}=2\omega$ for a wide range of $A$.
Two cross-sections with $A = \omega$ and $8\omega$ are presented in Fig.~\ref{fig-CPT} (b) and (c), respectively.  These panels show that with increasingly strong driving $A/\omega$, the contribution of the off-diagonal coupling $\xi$ becomes less important, and $|\psi(\pi/4) \rangle$ becomes ``darker''.
For the state orthogonal to $|\psi(\pi/4) \rangle$, with $\tan(\theta)=-1$, a resonant peak appears as is shown in both Fig.~\ref{fig-CPT} (b) and (c).

\subsection{Quantum Gates via the Driving-Induced Synthetic Coupling}\label{sec-qgate}

As shown in the effective Hamiltonian (\ref{eq-H_eff}), the higher order interference produces a synthetic coupling $\xi$ [see Eq.~(\ref{eq-Xi})], which could be employed to generate quantum gates, and has the potential for fast operation.
Here we demonstrate qubit control in a five-level ST system.
As we have discussed in the Appendix, though the relevant $H_\text{eff}$ in this case is in principle for five states, nearby a resonance only three states are relevant as long as the difference between resonant frequencies is significantly greater than the synthetic couplings. Thus we can continue to use Eq.~(\ref{eq-H_eff}) as our starting point.

Different from the previous section, here we do need a precise driving amplitude, which ensures suppression of leakage current caused by $g_{1,2}J_\nu$, while generating an accurate value for $\xi$.  According to Eq.~\ref{eq-H_eff}, leakage to the shuttle state is negligible as long as $|\delta|\gg |g_i J_\nu|$, while the spin manipulation time is determined by the strength of the synthetic coupling $\xi$.
To suppress leakage, one can work in the large detuning regime \cite{Zajac2018, Takeda2020, Hendrickx2020}, leading to a higher order of $J_\nu$ in $H_\text{eff}$.  Unfortunately large detuning also suppresses $\xi$, and is thus not a useful regime for quantum gates.
On the other hand, by tuning the driving amplitude $A/\omega$, leakage can also be suppressed while maintaining a finite value for $\xi$.
In other words, it is possible to implement state manipulation in the small-detuning regime, which has the additional benefit of a greater exchange coupling strength in the ST system.

In Fig.~\ref{fig-Rabi}(a) we present the transition amplitude between $|\uparrow, \uparrow\rangle$ $\leftrightarrow$ $|\uparrow, \downarrow\rangle$
($T_{12}$) and $|\downarrow, \downarrow\rangle$ $\leftrightarrow$ $|\downarrow, \uparrow\rangle$ ($T_{34}$) versus frequency and detuning energy $\varepsilon$.
At the point marked by the black triangle near zero detuning, these two transitions are degenerate, which means that the spin rotations in the two dots are independent of each other, and single-spin rotation in each dot can be generated.  Otherwise, if transitions in one dot occur at different driving frequencies dependent upon the spin orientation in the other dot, spin rotation in one dot would be conditional on the spin state in the other dot, leading to controlled rotation and two-spin gates.

In Fig.~\ref{fig-Rabi} (b) and (c), we present the time evolution of transition probability $P_{ij}(t) = |\langle i|U(t)|j\rangle|^2$, where $U(t)$ is the time evolution operator in the original frame.
Fig.~\ref{fig-Rabi} (b) shows single-spin rotation with parameters $g_{1}=1$ Gc, $g_2 = 2$ Gc and a strong driving amplitude of $A = 86.4$ Gc.
The cross-section of the chevron at resonance exhibits high-fidelity exceeding 99.9\% with fast manipulation time of $35$ ns.
When $\varepsilon \ne 0$, the two transitions shift differently as shown in Fig.~\ref{fig-Rabi} (a), and a controlled rotation can be generated.
In Fig.~\ref{fig-Rabi} (c), we present controlled rotation with $\varepsilon\approx 26$ Gc and $A=20$ Gc, as indicated by the red star in Fig.~\ref{fig-Rabi}(a).
The result contains two overlapping chevrons --- one is narrowly peaked at $\omega - D_{12} \approx -0.2$ Gc and one is broadly peaked at $\omega - D_{12} \approx 0.2$ Gc. They are the consequence of $T_{34}$ and $T_{12}$, respectively.
The cross-section indicated by the black dashed line in Fig.~\ref{fig-Rabi} (c) shows a resonant transition with amplitude over $99.5\%$ and a suppressed off-resonant rotation amplitude of $0.4\%$.
The frequency difference between the two resonant peaks indicates a significant exchange coupling, which is a result of the small detuning.

\begin{figure}
    \centering
    \includegraphics[width=8.3 cm]{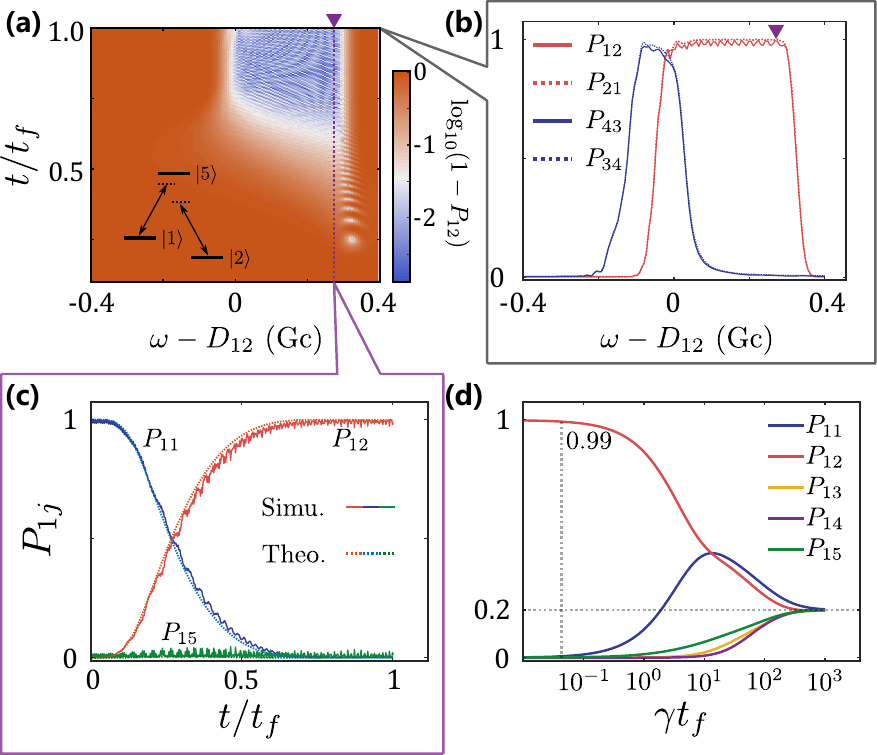}
	\caption{{\bf Quantum gate via an adiabatic passage in the ST system}. (a) Logarithm of infidelity $\log_{10}(1-P_{12}(t))$ with respect to frequency derivation $\omega - D_{12}$.
	Inset shows the  effective three-level model by $|1\rangle$, $|2\rangle$, and $|5\rangle$.
	(b) The final probability $P_{ij}(t_f)$ as a function of $\omega-D_{12}$, in which the triangle point can be used as the working point of controlled-rotation gate.
	(c) The cross-section along the dashed line in (a), from both full simulation of the time-dependent Hamiltonian (solid curves) and that by the three-level effective Hamiltonian $H_\text{eff}$ (dashed curves).
    (d) Probability with Lindblad dissipation strength $\gamma$.
    For $t_f < T_2/10$, transition amplitude $P_{12}>99\%$; and for $t_f=T_2/20$ and in the long-time limit, state thermalization is reached, with $P_{1i} = 1/5$ when $t_f > 10^2T_2$. }
	\label{fig-adiabatic}
\end{figure}

\subsection{Adiabatic state transfer by driving amplitude modulation} 
\label{sec-AST}

In an atom describable by a $\Lambda$ model in quantum optics, adiabatic state transfer can be achieved via modulation of the coherent transverse couplings \cite{Kuklinski1989, scully1997, Vitanov2017}.  Such state transfer has been demonstrated in various systems beyond three-level atoms \cite{Takekoshi2014, Kumar2016}, including in quantum dot spin chains in a mathematically isomorphic process called adiabatic quantum teleportation (AQT) \cite{Oh2013, Kandel2021}.
In our longitudinally driven system, to achieve state transfer we need to adiabatically modulate the effective Hamiltonian $H_{\text{eff}}(t)$, assuming that the modulation frequency is much smaller than the driving frequency.
Compared to previous works where the off-diagonal coupling terms are tuned independently, $g_i$ here are assumed to be constants in our model, leaving only the slow-varying amplitude $A(t)$ to be the control parameter.

In the following example, we accomplish adiabatic state transfer in a five-level ST system by modulating the driving amplitude as
\begin{equation}
    A(t) = 100\sin^2(\pi t/2t_f), \quad 0 \le t \le t_f,
\end{equation}
where $t_f = 1$ $\mu$s is the total evolution time. This choice of $A(t)$ is for illustration purpose only and is not optimized.  In the effective Hamiltonian, all the correlations such as $\xi$, $\Delta$, and $\Delta'$ are included because when $A(t)$ is small, they may be important.
We investigate the adiabatic process both through the effective Hamiltonian and by numerically solving the time-dependent Schr\"odinger equation with different driving frequencies.  As we have discussed previously, the effective model Hamiltonian here is equivalent to that for a three-level $\Lambda$ configuration composed of $|1\rangle$, $|2\rangle$ and $|5\rangle$, while the effective couplings are only tuned by driving amplitude via the Bessel functions.

With the modulation on the driving amplitude, state transfer occurs in a wide range of driving frequencies as shown in Fig.~\ref{fig-adiabatic} (a), where we have plotted the logarithm of infidelity $1-P_{12}(t)$.
In Fig.~\ref{fig-adiabatic} (b) we present different transition probabilities $P_{ij}(t_f)$ with the same adiabatic passage as a function of the driving frequency.
Based on the fact that transitions are addressable via driving frequency, controlled rotation can be constructed.
For example, when $\omega-D_{12}=0.27$ Gc as is labeled by the solid purple triangle in both Fig.~\ref{fig-adiabatic} (a) and (b), $P_{12}$ and $P_{21}$ are significant while $P_{34}$ $P_{43}$ are minimal.
In addition, the plateau of $P_{ij}(t_f)$ indicates that the adiabatic state transfer is robust to frequency/energy spacing shift.
In Fig.~\ref{fig-adiabatic} (c), we compare the time evolution from our full numerical simulation with the results from $H_\text{eff}(t)$ for a given driving frequency $\omega$, along the dashed vertical line in Fig.~\ref{fig-adiabatic} (a).  We find good agreements between the two approaches, which also clearly illustrates that the three-level model is sufficient to capture the dynamics of the five-level system in our discussion.
Furthermore, in Appenidx \ref{appdx-comp}, we show that $H_{\text{eff}}$ in Eq.~\ref{eq-H_eff} is a more robust and faithful representation of the original time-dependent Hamiltonian than that from the high-frequency expansion \cite{Goldman2014}.

In a semiconductor DQD, dephasing is usually quite fast, especially among states with different charge distributions.  We thus also investigated its influence on our adiabatic state transfer protocol, described by the following Lindblad master equation
\begin{equation}
    \dot\rho = -i[H(t), \rho] + \sum_{k\ne 0} \left[L_k\rho L_k^\dagger-\frac{1}{2}\{L_k^\dagger L_k, \rho\}\right],
\end{equation}
where the summation represents non-unitary evolution and $L_k = \sqrt{\nu_k}|k\rangle\langle k|$, for $k=1,\ldots, 5$, account for the white-noise energy fluctuations for each level.  We have neglected the contribution of relaxation among these levels for the reason that in QDs, $T_1 \gg T_2$ in general.
In Fig.~\ref{fig-adiabatic} (d) we plot $P_{1j}(t_f)$ as a function of $\gamma$ (assuming $\nu_k = \gamma = 1/T_2$),
showing a transition from the perfect state transfer with $P_{12}(t_f)\rightarrow 1$ to $P_{1j} = 1/5$ as the system dephase into a completely mixed state, which is reached when $t_f > 10^2 T_2$. If the protocol time is much shorter than dephasing time, for example when $t_f = T_2/20$, the fidelity $P_{12}(t_f)$ can exceed 0.99, as expected.

\section{Conclusion}

In conclusion, we have demonstrated theoretically full tunability for a multilevel quantum system by introducing a longitudinal drive.  Specifically, we develop a universal approach that allows us to study the quantum coherent dynamics of a multilevel system under longitudinal drive. We show that the resulting multilevel dynamics is well described by an effective three-level model, and is fully-tunable by the driving field.
Employing this effective model, we demonstrate a wide variety of quantum coherent phenomena that are qualitatively similar to driving induced splitting (or ATS) and dark state in quantum optics.  We show that longitudinal drive leads to some distinct features in the spectrum and dynamics as compared to transversely driven systems.  For example, we are able to understand the so-called odd-even effect caused by coherent destruction of tunneling within our effective model.
We also demonstrate fast quantum gates based on the interference-induced synthetic coupling in a five-level ST system, and show that qubit manipulation with a broadband driving can be realized by adiabatic passage.

Our results clearly demonstrate that coherent longitudinal driving and the resulting multi-level interference lead to strong tunability, since coherent phases in the dynamical processes are controlled by the driving field.
In the context of quantum dot devices, such convenient tunability through driving of interdot detuning would be readily available for hybrid qubits and singlet-triplet qubits, and can be employed in any multi-dot device.
Furthermore, though we focus on a sinusoidal driving field in this initial study, our approach can be extended to general periodic driving field after a Fourier expansion.
Most importantly, the theoretical considerations we present here are generic and should be valid for a variety of multilevel systems that are experimentally relevant.

\section{Acknowledgments}
This work was supported by the National Natural Science Foundation of China (Grants No. 12074368, 92165207, 12034018 and 61922074), the Innovation Program for Quantum Science and Technology (Grant No. 2021ZD0302300), the Anhui Province Natural Science Foundation (Grants No. 2108085J03), the USTC Tang Scholarship.  
X.H. acknowledges financial support by U.S. ARO through grant W911NF1710257.

\appendix

\section{Derivation of the effective Hamiltonian} \label{appdx-Heff}

To solve the driven dynamics of a quantum system, a common practice is to apply the Floquet theorem and obtain an effective Hamiltonian that is time-independent. For example, following Ref.~\onlinecite{Goldman2014}, such an effective Hamiltonian can be obtained using a high-frequency expansion.  Here we present our approach in arriving at a time-independent effective Hamiltonian based on a more nuanced treatment of the spectrum.

\subsection{Rotating frame for longitudinally driven system}
We are dealing with a time-periodic system with longitudinal driving up to the strong driving limit ($A \gg \omega$).  The time dependence is contained in a diagonal element of the Hamiltonian, which cannot be treated conveniently using Floquet theorem in our searching of a time-independent $H_\text{eff}$.  We thus first perform a unitary rotation of $H$ by
\begin{equation}\label{eq-rotat_frame}
    \mathcal{U} = \text{diag}(e^{i(D_2 + n\omega) t}, e^{iD_2 t }, e^{i(D_2 + N\omega)t-i\frac{ A \sin (t \omega)}{\omega}}).
\end{equation}
The resulting Hamiltonian takes the form
\begin{equation}
    \tilde H(t)=
    \begin{pmatrix}
        D_{12}-n\omega  & 0 & c(t) g_1 e^{in\omega t}\\
        0 &    0& c(t) g_2 \\
        c^*(t) g_1 e^{-in\omega t} &c^*(t) g_2  & D_{32}-N\omega \\
       \end{pmatrix},
	\label{eq-Ht}
\end{equation}
with $c(t) = e^{\frac{i A \sin ( \omega t)}{\omega }-iN\omega t}=\sum_\nu J_\nu(A/\omega)e^{i(\nu-N)\omega t}$.
By ``proper'', we mean that (i) the diagonal matrix elements should be restricted to the first Brillouin zone (keeping in mind that we will perform Floquet expansion next); (ii) the new Hamiltonian $\tilde{H}(t)$ should have the same periodicity as $H(t)$, and (iii) the degrees of freedom in $U$ should be uniquely determined.  The time dependence of $\tilde H(t)$ is now conveniently located in the off-diagonal terms, which allows us to apply the Floquet theorem to treat the problem.

\subsection{Floquet-Schr\"odinger equation and numerical simulation.}

The Floquet theorem provides a general framework to study linear models with periodic modulation in the time domain \cite{Shirley1965}, and has also been exploited to study dissipative systems \cite{Kohler1997}.  Here we use Floquet theorem to derive a time-independent effective Hamiltonian for our driven system.

\begin{figure}
    \centering
    \includegraphics[width=8.3 cm]{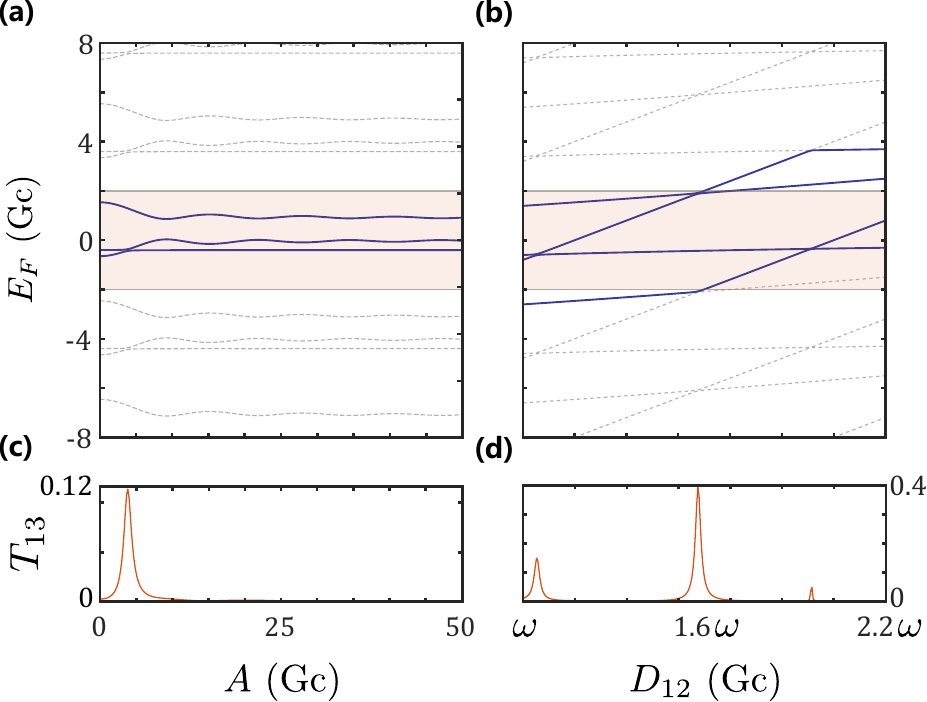}
    \caption{{\bf Floquet energy levels $\bm{E_F}$ and transition amplitude $\bm{T_{13}}$ versus $\bm A$ in (a) and $\bm {D_{12}}$ in (b), respectively.} 
    The energy levels are presented in solid blue or dashed gray lines, while the red shading area indicates the first Brillouin zone. 
    The energy structure shows a periodicity as Floquet theorem indicated.
    When energy levels crossover with each other, state transfer occurs, thus population in $|3\rangle$ can be detected as indicated by solid red lines in (c) and (d).}
    \label{fig-Floquet_levels}
\end{figure}

For $i\partial_t|\psi\rangle = \tilde H(t) |\psi\rangle$, where $\tilde H(t)$ was defined above, we can write $|\psi\rangle = e^{-iqt} |u(t)\rangle$, where $q$ is the quasi-energy and $|u(t)\rangle$ is a periodic function with period $T$.
We now have
\begin{equation}
	(\tilde H(t) - i\partial_t)|u(t)\rangle = q |u(t)\rangle, \quad |u(t)\rangle = |u(t + T)\rangle.
\end{equation}
This equation can be cast into a stationary equation, assuming that $|u(t)\rangle = \sum_{\alpha, n} c_{n\alpha}|\alpha, n\rangle$,
where $|\alpha, n \rangle = e^{in\omega t} |\alpha \rangle$.
If one takes $\left(\mathscr{U}\right)$ as a column matrix consisting of all the $c_{n\alpha}$ in proper order with $n \in \mathbb{Z}$, we find the Floquet-Schr\"odinger equation in the matrix form as $\mathscr{H}_F \left(\mathscr{U}\right) = q \left(\mathscr{U}\right)$, with matrix elements
\begin{equation}
    \langle \alpha,n|\mathscr{H}_F|\beta, m\rangle = {\tilde H}_{\alpha\beta}^{(n-m)}+n\omega\delta_{nm}\delta_{\alpha\beta},
    \label{eq:Floquet_mat_elemt}
\end{equation}
where ${\tilde H}^{(n-m)}_{\alpha\beta}$ is the $(n-m)$th Fourier coefficient of matrix element ${\tilde H}_{\alpha\beta}$.
Though $\mathscr{H}_F$ is a matrix of infinite size, in numerical simulations we can make a truncation for $n$ (in our case we choose $|n| \le N_c = 20$) to calculate the quasienergy $q$, which is shown in Fig.~\ref{fig-Floquet_levels} (a)-(b).
As expected, the spectrum is a repetition of levels in the first Brillouin zone shifted by $n\omega$.  Physically, the dynamics in the higher Brillouin zones correspond to the results from the higher harmonics of the driving field.  Thus we will in general focus on the first Brillouin zone defined as $\mathscr{B} =[-\omega/2, \omega/2]$.

Based on $\mathscr H_{F}$, we can evaluate the transition amplitude from $|\beta\rangle$ to $|\alpha\rangle$ by
\begin{equation}
    U_{\alpha,\beta}(t;t_0)=\sum_n\langle\alpha,n|e^{-i\mathscr H_{F}(t-t_0)}|\beta,0\rangle e^{in\omega t},
\end{equation}
where the summation collects contributions from all orders of Fourier components. The averaged transition amplitude is then given by
\begin{align}
    T_{\alpha\beta}(t_0) &= \lim_{t\rightarrow\infty} \frac{1}{t} \int_0^t |U_{\alpha,\beta}(\tau;t_0)|^2 d\tau\\
     &= \sum_{m,n, \phi} \langle\alpha,n|\phi\rangle\langle \phi|\beta,0\rangle \langle\beta,m|\phi\rangle\langle \phi|\alpha,n\rangle e^{im\omega t_0},
\end{align}
with $|\phi\rangle$ the eigenstates of $\mathscr H_{F}$.

Our numerical results for this transition amplitude are shown in Fig.~\ref{fig-Floquet_levels} (c) and (d), where resonance transitions are expected at degeneracy points.

\subsection{Effective Hamiltonian} 

Finally, we obtain a time-independent effective Hamiltonian $H_\text{eff}$ by eliminating the off-diagonal couplings between different Brillouin zones in $\mathscr H_F$ to the lowest order.  This is achieved by L\"{o}wdin's quasi-degenerate perturbation theory via a unitary transformation [equivalent to the Schrieffer-Wolff (SW) transformation] \cite{Winkler2003} of $\tilde{\mathscr H}_{F} = U^\dagger \mathscr H_{F} U$, where $U = e^{-S}$, and $S$ is given by
\begin{equation}
    S_{ml} =  \frac{({\mathscr H_F})_{ml}}{({\mathscr H_F})_{ll}-({\mathscr H_F})_{mm}} \,.
    \label{eq-Sml}
\end{equation}
Here $m$ and $l$ include indices $n$ and $\alpha$ in Eq.~\ref{eq:Floquet_mat_elemt}.

The effective Hamiltonian $H_{\rm eff}$ in the main text is thus obtained, in which
$\Delta', \Delta, \xi$ are the lowest order corrections from the higher Brillouin zones:
\begin{eqnarray}
	&& \Delta' = \sum_{m \ne N-n}\frac{(g_1 J_m)^2}{D_{13}+m\omega}, \quad \Delta = \sum_{m \ne N}\frac{(g_2 J_m)^2}{D_{23}+m\omega}, \label{eq-Deltaprime}\\
	&& \xi = \sum_{m\ne N}\frac{g_1g_2J_mJ_{m-n}}{2(D_{23}+m\omega)} + \sum_{m\ne N-n}\frac{g_1g_2J_mJ_{m+n}}{2(D_{13}+m\omega)}.
	\label{eq-Xi}
\end{eqnarray}

Notice that the above equations do not exclude the resonance given in Eq.~\ref{eq-resonance}. SW transformation accounts for the influence from sectors other than FBZ (ensured by the summation range), while the resonance condition is set within FBZ.

\section{Three-level model versus five-level model}\label{appdx-5vs3}
In the main text, we use a three-level effective model to describe the driven five-level system near certain resonances.
The numerical simulation based on the time-dependent five-level Hamiltonian in Fig.~\ref{fig-adiabatic} agrees well with the theoretical result based on a three-level model, indicating that such an approximation is indeed valid.
Here we discuss under what circumstances can a driven five-level system be reduced to an effective three-level one. 

The driven five-level system is governed by the following Hamiltonian
\begin{equation}
    H_5(t) = 
    \begin{pmatrix}
        D_1 & 0 & 0 & 0 & g_1\\
        0 & D_2 & 0 & 0 & g_2\\
        0 & 0 & D_3 & 0 & -g_2\\
        0 & 0 & 0 & D_4 & g_1\\
        g_1 & g_2 & -g_2 & g_1 & D_5(t)
    \end{pmatrix},
\end{equation}
where $D_5(t) = \varepsilon - A\cos(\omega t)$. Following the same procedure as presented in section \ref{appdx-Heff}, we first perform a rotation by 
\begin{equation}
    \mathcal U = \text{diag}(1, 1, 1, 1, e^{-i\frac{A\sin(\omega t)}{\omega}}).
\end{equation}
For illustration purpose, we do not introduce offsets as we did in Eq.~\ref{eq-rotat_frame}. 
The time dependence is now in the off-diagonal elements, and the resulting Hamiltonian is given by
\begin{equation}
    \tilde H_5(t) = 
    \begin{pmatrix}
        D_1 & 0 & 0 & 0 & \tilde g_1(t)\\
        0 & D_2 & 0 & 0 & \tilde g_2(t)\\
        0 & 0 & D_3 & 0 & -\tilde g_2(t)\\
        0 & 0 & 0 & D_4 & \tilde g_1(t)\\
        \tilde g_1^*(t) & \tilde g_2^*(t) & -\tilde g_2^*(t) & \tilde g_1^*(t) & \varepsilon
    \end{pmatrix},
\end{equation}
where $\tilde g_i(t) = g_i\sum_\nu J_\nu(A/\omega)e^{i\nu \omega t}$.
This result indicates that when we focus on the resonance condition, e.g., $D_1-D_2 = n\omega$, we can eliminate $|3\rangle$ and $|4\rangle$ as long as the energy differences are significant compared to coupling strengths $g_i$.
Under this condition, the dynamics of interest, such as state transfer between $|1\rangle$ and $|2\rangle$, can be captured by a three-level model spanned on $|1\rangle$, $|2\rangle$ and $|5\rangle$, with $|5\rangle$ mediating the lowest order processes between $|1\rangle$ and $|2\rangle$, while $|3\rangle$ and $|4\rangle$ only contributing through higher-order off-resonance processes.

As mentioned in the main text, we set $g_1 = 1$ Gc, $g_2 = 2$ Gc, $D_1 = -D_4 =  40$ Gc and $D_2 = -D_3 =  10$ Gc when performing the simulation in Fig.~\ref{fig-Rabi} and Fig.~\ref{fig-adiabatic}.
These parameters ensure the accuracy of the three-level model, and agreement between theoretical results based on the effective Hamiltonian and the numerical simulation based on the full Hamiltonian.

\section{Comparison of different theoretical approaches} \label{appdx-comp}

\begin{figure*}
    \centering
    \includegraphics[width=12.3 cm]{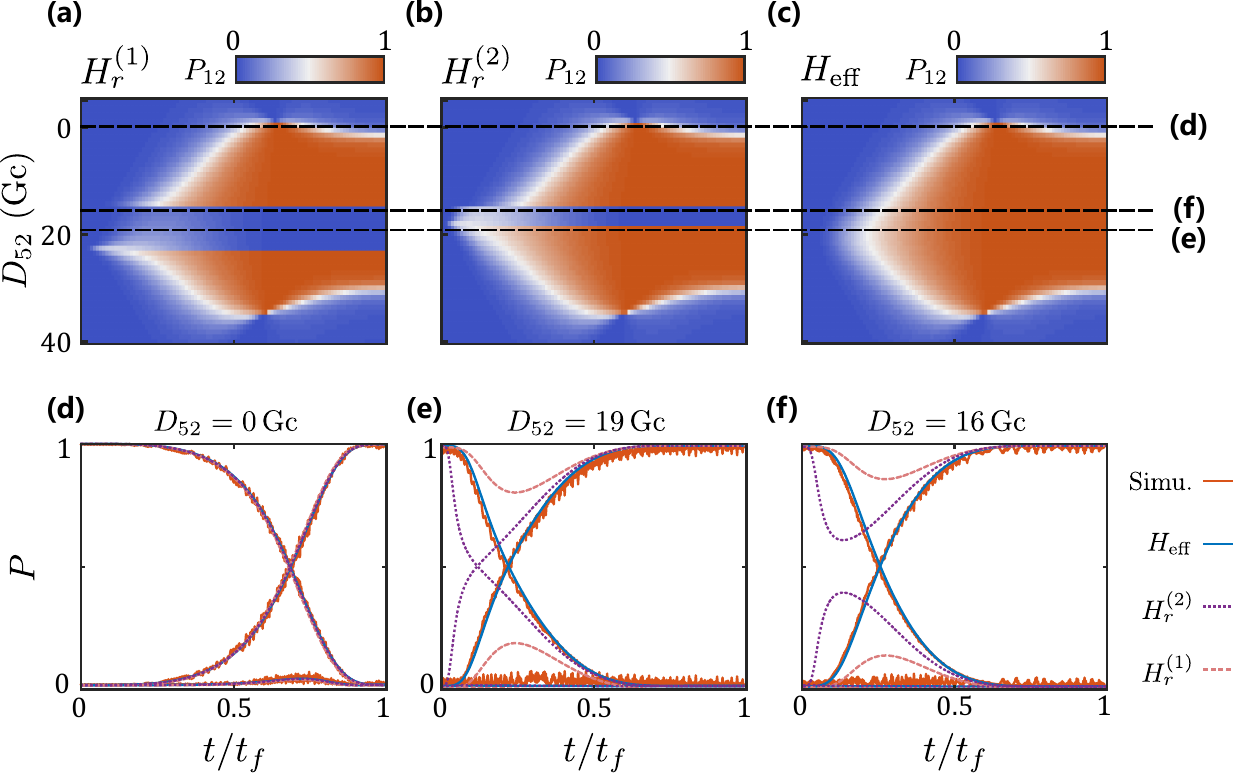}
    \caption{
    {\bf Comparison of two effective Hamiltonians.}
    Adiabatic state transfer is investigated to clarify the difference between two models.
    We simulate the ideal adiabatic state transfer as a function of detuning $D_{52}$ based on three different models -- first-order $H_r^{(1)}$ (a), second-order $H_r^{(2)}$ (b), and our effective model $H_\text{eff}$ (c). 
    For clarity, we present three cross-sections labeled by dashed lines in (d), (e), and (f).
    $H_\text{eff}$ faithfully represents the dynamics in (d), (e) and (f), while $H_r^{(1,2)}$ only accurate in (d).
    }
    \label{fig-Heff_comparison}
\end{figure*}
 
The objective in deriving an effective model is to find a basis in which the originally time-dependent Hamiltonian becomes approximately time-independent, so that the associated eigenstates are stationary. This approach is aptly demonstrated in going to the rotating frame when treating conventional nuclear magnetic resonance.  In Ref.~\onlinecite{Goldman2014}, a universal effective model for a driven system was obtained based on high-frequency expansion.  Here we compare our effective model with this established model, and show that the basis found in our approach is closer to reaching this objective, such that dynamics based on our effective Hamiltonian is more consistent with numerical simulations under wider range of conditions.

According to Ref.~\onlinecite{Goldman2014}, one can find a rotation $e^{-iK(t)}$ such that $H_r = e^{-iK(t)}H(t)e^{iK(t)}$ is time-independent.
Following the same rotation defined by $\mathcal U$ above, one can write $\tilde H(t) = H_0 + \sum_{n\ne 0} H_n e^{in\omega t}$, and the effective Hamiltonian $H_r$ is obtained as 
\begin{equation}
	H_r = H_0 + \frac{1}{2}\sum_{n \ne 0} {\frac{[H_n, H_{-n}]}{n\omega}} + \cdots.
\end{equation}

The $H_r$ thus obtained is equivalent to our $H_{\text{eff}}$ when $D_{12}$ and $D_{52}$ are integer multiples of $\omega$, but is otherwise different.
To demonstrate their differences quantitatively, we repeat the numerical simulations as presented in Fig. 5 (c).  Here we keep $D_{12} \approx \omega$, while varying $D_{52}$, and compare results given by the two models and the original time-dependent Hamiltonian.  
For clarity, we denote the results derived by Ref.~\onlinecite{Goldman2014} as $H_r^{(1)}$ and $H_r^{(2)}$ for first- and second-order approximation, respectively.  

In Fig.~\ref{fig-Heff_comparison} (a), (b), and (c) we track the transfer probability $P_{12}$ as a function of $D_{52}$ and evolution time $t$.
A gap appears in both panels (a) and (b), indicating that no transfer could occur when $D_{52}$ is in those ranges, while there is no gap in the results obtained from $H_\text{eff}$ in (c).  Furthermore, the gap in the results from $H_r^{(2)}$ is narrower than that from $H_r^{(1)}$.  In Fig.~\ref{fig-Heff_comparison} (d), (e), and (f) we choose three cross-sections denoted by dashed lines in (a), (b), and (c) to demonstrate how the different models differ.  For $D_{52}=0$, presented in panel (d), $H_r^{(1)}$ is equivalent to $H_\text{eff}$, while $H_r^{(2)}$ gives no correction to $H_r^{1}$.  All three models are consistent with the direct numerical simulation.  However, when $D_{52}\ne 0$, difference arises between $H_r^{(1,2)}$ and $H_\text{eff}$.  In Fig.~\ref{fig-Heff_comparison} (e), only $H_\text{eff}$ is quantitatively consistent with the numerical simulation, while $H_r^{(2)}$ is only qualitatively consistent.  The results from $H_r^{(1)}$, on the other hand, are qualitatively different from the simulation results, with no population transfer at the long time limit.  When $D_{52}=16$ Gc, as presented in Fig.~\ref{fig-Heff_comparison} (f), both $H_r^{(1,2)}$ fail to capture the adiabatic state transfer process, while $H_\text{eff}$ still works well and track the numerical simulation perfectly.

To further accentuate the differences between the two effective Hamiltonians, below we rederive our $H_\text{eff}$ using a procedure similar to that in Ref.~\onlinecite{Goldman2014}.
Different from Ref.~\onlinecite{Goldman2014}, here we prove that for any $e^{iqt}$ ($q$ being an operator, thus a matrix when the basis states are chosen), there exists an operator $K(t)$ and a rotation $e^{-iK(t)}$, such that $e^{iK(t)}e^{iqt}H(t)e^{-iqt}e^{-iK(t)} \approx e^{iqt} H_\text{eff} e^{-iqt}$, where $H_\text{eff}$ is time-independent.
For convenience, we denote 
\begin{align}
    H'(t) &= e^{iqt}H(t)e^{-iqt}\\
        &= e^{iqt}\left(H_0+\sum_{n\ne0}H_ne^{in\omega t}\right)e^{-iqt}\\
        &= H_0'+\sum_{n\ne 0}H_n'e^{in\omega t },
\end{align}
with $H_{0,(n)}'=e^{iqt}H_{0,(n)}e^{-iqt}$.
The objective now is to find a $K(t)$ such that $e^{iK(t)}H'(t)e^{-iK(t)}$ is independent of $\omega$ rather than $t$.
In this way, we find that
\begin{equation}\label{eq-Kt}
    \partial_tK\approx\sum_{n\ne0}H_n'e^{in\omega t}.
\end{equation}
If we further suppose $q$ is diagonal as $q=\text{diag}\{q_1,\ldots,q_n\}$, we obtain
\begin{equation}
    H_\text{eff}\approx H_0+\frac{1}{2}\sum_{n\ne0}[Q_n,H_{-n}],
\end{equation}
where
\begin{equation}
    Q_n=\sum_{j,j'}\frac{(H_n)_{jj'}}{n\omega+q_j-q_{j'}}|j\rangle\langle j'|.
\end{equation}
Recall that we introduced two energy offsets $n\omega$, $N\omega$ in Eq.~\ref{eq-rotat_frame},
thus, if $q_i$ are set as $D_i-m\omega$, where $m\omega$ corresponds to energy offsets, i.e., $q = \text{diag}\{D_1+n\omega, D_2, \varepsilon+N\omega\}$, one obtains $H_{\rm eff}$ presented in the main text. 
On the other hand, if $q$ is set as the identity matrix, one obtains $H_r$.
Notice that $H_r$ and $H_\text{eff}$ are only different in the denominator of $Q_n$. As a consequence, $H_r$ can be equivalent to $H_\text{eff}$ as we mentioned before when $D_{12}=n\omega$, $D_{32} = N\omega$.
In addition, the new basis is given by 
\begin{equation}
    |\psi_\text{eff}\rangle = e^{-iqt}e^{-iK(t)}|\psi\rangle,
\end{equation}
with $|\psi\rangle$ the basis in the original frame and
\begin{equation}
    K\approx\int_0^t\sum_{n\ne 0}H_n'(\tau)e^{in\omega \tau}d\tau + \mathcal{O}(1/\omega^2).
\end{equation}

Both $H_\text{eff}$ and $H_r$ are time-independent.  However, from the perspective of perturbation theory, $H_r$ does not take into account the energy difference $D_i-D_j$, and is thus not as accurate as $H_\text{eff}$ whenever these energy differences are significant. We can introduce energy offsets to ease this conflict, so that the difference between $H_r$ and $H_\text{eff}$ mainly arises from $\delta D_\omega = mod(D_i-D_j, \omega)$, with $mod()$ the modulo operation.
Hence, whenever $n\omega \gg |\delta D_\omega|$, $H_r$ is approximately equivalent to $H_\text{eff}$.  In contrast, when $|\delta D_\omega|$ is comparable to $n\omega$, $H_r$ is no longer accurate.
For example, the maximal deviation can be seen when $D_{52}=\omega/2$, as shown in Fig.~\ref{fig-Heff_comparison}, where $\omega/2 = 15$ Gc.

In short, the analytical analysis presented here and comparisons with numerical simulations show that our effective model $H_{\rm eff}$ is a faithful representation of the dynamics of the longitudinally driven multi-level system.  On the other hand, while the high-frequency expansion is an elegant theoretical approach, the resulting effective Hamiltonian often fails to catch the important features of the original driven system, particularly in the strong-driving limit.

\section{Driving induced resonance shifts and odd-even effect.} \label{appdx-odd-even}

\begin{figure}[]
	\centering
	\includegraphics[width=8.3cm]{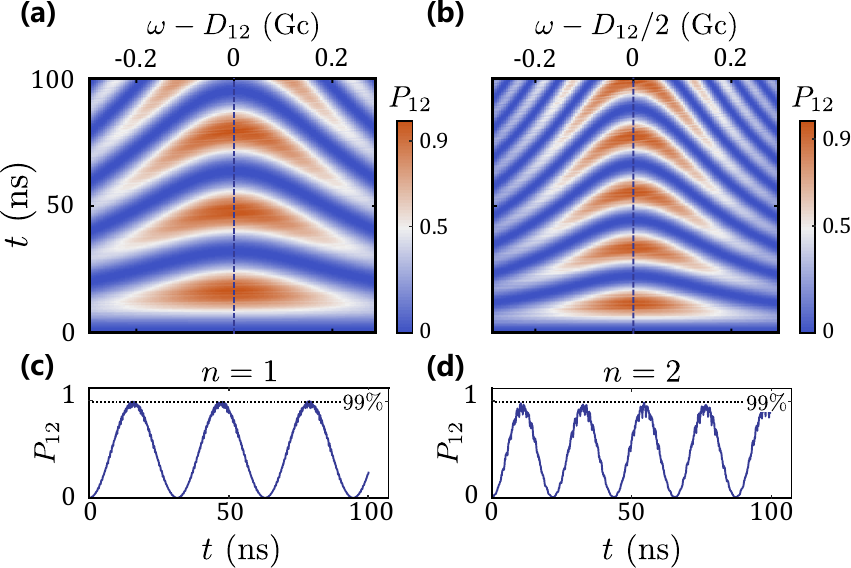}
	\caption{{\bf Rabi oscillation based on higher harmonic resonance.} Rabi oscillation chevrons for $n=1$ (a) and $n=2$ (b), resonant oscillation amplitude of which is presented in (c) and (d) respectively.}
	\label{fig-multifreq}
\end{figure}

To better understand the resonant transitions presented in the main text, we first diagonalize the subspace of $\{|2\rangle_\text{eff}, |3\rangle_\text{eff}\}$ into $\{|+\rangle, |-\rangle\}$. 
The effective Hamiltonian can then be written as 
\begin{equation}\label{eq:H_eff_sub_diag}
    H_{\rm{eff}} = 
    \begin{pmatrix}
        D_{12} - n \omega & r_+ & r_- \\
        r_+^* 	& E_+ & 0 \\
        r_-^* 	& 0 & E_-
    \end{pmatrix},
\end{equation}
with $E_\pm$ the eigenenergies of subspace spanned by $\{|2\rangle_\text{eff}, |3\rangle_\text{eff}\}$, and $r_\pm$ the corresponding coupling strength to $|1\rangle_\text{eff}$.
For weak couplings, the resonant condition is determined by $D_{12} - n\omega= E_\pm$, resulting in the pair of absorption peak induced by longitudinal drive as mentioned in the main text.
Notice that the resonant condition implies that state manipulation can be achieved by higher harmonics. 
To demonstrate this point, we present a numerical simulation result in Fig.~\ref{fig-multifreq} (a), (b), where we present Rabi oscillation for $D_{12}=\omega$ and $D_{12}=2\omega$.

With $r_\pm$ modulated by the Bessel functions, the effective couplings between $|\pm\rangle$ and $|1\rangle_\text{eff}$ vanishes for certain values of the driving amplitude $A$, in analogy to the so-called coherent destruction of tunneling in two-level system \cite{Grossmann1991, Oliver2005, Ashhab2007, Stehlik2012}.  Furthermore, the presence of the Bessel functions also induces a parity-dependent coupling, which we refer to as odd-even effect in the main text.
According to $H_\text{eff}$, the resonances occur at $D_{12}-n\omega = E_\pm$.
One of the branches, with $D_{12}-n\omega\approx 0$, gives rise to the parity effect.
When $D_{12}=n\omega$, we have 
\begin{equation}
    \begin{aligned}
        r_- &\propto \xi g_2J_{N} - g_1J_{N-n} \Delta\\
        &= \sum_{m\ne N}\frac{g_1g_2^2}{m\omega + D_{23}}(J_{m-n}J_{m}J_{N}-J_{m}^2J_{N-n}),
    \end{aligned}
\end{equation}
on the other hand, for the other branch, in the vicinity of $D_{13}=(N-n)\omega$, we have
\begin{equation}
    \begin{aligned}
        r_+ &\propto\sum_{m\ne N}\frac{g_1g_2^2J_mJ_{m-n}J_N}{2(D_{23}+m\omega)}+\sum_{m\ne N-n}\frac{g_1g_2^2J_mJ_{m+n}J_N}{2(m+n-N)\omega}\\
        &+g_1J_N\left( D_{32}-N\omega-\sum_{m\ne N}\frac{(g_2J_m)^2}{D_{23}+N\omega} \right)
    \end{aligned}
\end{equation}
For Bessel functions of the first kind, we know the asymptotic behavior of
\begin{equation}
    J_n(x)\approx \sqrt{\frac{2}{\pi x}}\cos[x-(2n+1)\pi/4],~x \gg n.
\end{equation}
This guarantees $r_-\approx 0$ at $D_{12}=n\omega$ for even $n$s when $A/\omega \gg N-n,~m-n$, while for odd $n$s, zero points can be found around $D_{12}=n\omega$ when $J_{m-n}J_{N}= J_mJ_{N-n}$, but this condition could not always be fulfilled.
However, odd-even effect would disappear when $\varepsilon > A$, i.e., when system does not go through anti-crossings, which is in accordance with experimental findings \cite{Stehlik2014}.
It can be interpreted with the relation $N \approx D_{23}/\omega > A/\omega$, which means $\varepsilon>A$ undermines the approximation condition $A/\omega\gg N-n$ for small $n$.

\section{Odd-even effect with dissipation at the strong driving limit} \label{appdx-odd-even-diss}

In this section, by introducing dissipation and strong driving, we demonstrate how driving induced modulation of the resonances reported in the main text is in agreement with the odd-even effect reported by Ref.~\onlinecite{Stehlik2014}.

\begin{figure}[]
    \centering
    \includegraphics[width=8.6cm]{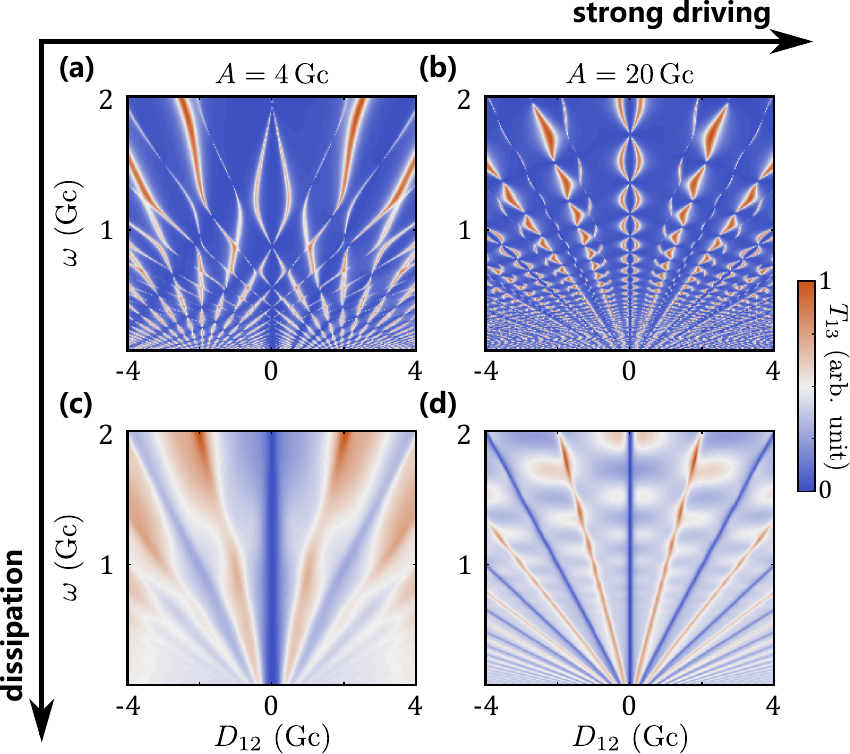}
    \caption{{\bf ATS signal versus $D_{12}$ and $\omega$ in different situations.}
    We compare ATS without dissipation under different driving amplitude, i.e., $4$ Gc in (a) and $20$ Gc in (b).
    We also repeat this comparison with dissipation, and the relevant results are presented in (c) and (d).
    }
    \label{fig-odd_even}
\end{figure}

Based on the fact that shuttle state $|3\rangle$ can be detected electrically, we calculate $P_{13}(t)$ by Lindblad master equation with and without dissipation under different driving amplitudes, and average $P_{13}(t)$ over $800$ ns to obtain $T_{13}$ approximately.
Dissipation is introduced through dephasing on each state at a rate of $\gamma=0.01$ GHz, and an electron jumping in and out of the DQD at a rate of $\Gamma=1$ GHz.
We calculate $T_{13}$ as a function of driving frequency $\omega$ and energy difference $D_{12}$, and the result is presented in Fig.~\ref{fig-odd_even}.

Figure \ref{fig-odd_even} (a) and (b) show resonance signals under a driving amplitude of $4$ Gc and $20$ Gc, respectively, and without dissipation.
As driving amplitude increases, the splitting of the resonance signal eventually saturates. As expected, when the driving field is sufficiently strong, resonant transition only happens at $D_{12}=n\omega$ or $D_{13}=m\omega$.
Furthermore, as discussed in the main text, for $D_{12}=2n\omega$, transition from $|1\rangle_\text{eff}$ to $|-\rangle$ is almost forbidden (determined by $r_-$).
From this perspective, even orders of harmonics are supposed to vanish in the strong driving limit, which yields the so-called odd-even effect.
We then take dissipation into account and repeat the simulation. The results are presented in Fig.~\ref{fig-odd_even} (c), (d), corresponding to the same parameters as in (a) and (b), respectively. The dissipation yields a background signal and broadens the resonant transition signal. On the basis of the background signal, we see harmonics of different orders, and the odd-even effect is quite apparent: even orders manifest as a dip meanwhile odd orders as a peak.
The results presented in Fig.~\ref{fig-odd_even} (d) almost reproduces the phenomenon in Ref.~\onlinecite{Stehlik2014} and the theoretical results in Ref.~\onlinecite{Danon2014}.
Comparing to the previous numerical works, we present a clear physical picture for the odd-even effect based on our effective Hamiltonian and the Bessel function modulated coupling matrix elements.







\bibliography{reference.bib}
\end{document}